\documentclass[3p]{elsarticle}

\usepackage{moreverb}

\usepackage[colorlinks,bookmarksopen,bookmarksnumbered,citecolor=red,urlcolor=red]{hyperref}

\usepackage{amsmath,amsfonts}
\usepackage{graphicx,float}
\usepackage{subfigure}

\newcommand{\barr}{\begin{array}}
\newcommand{\earr}{\end{array}}
\newcommand{\be}{\begin{equation}}
\newcommand{\ee}{\end{equation}}
\newcommand{\beqa}{\begin{eqnarray}}
\newcommand{\eeqa}{\end{eqnarray}}

\begin{document}
\begin{frontmatter}

\title{ Scattering of a short electromagnetic pulse 
from a Lorentz-Duffing film: theoretical and numerical analysis}

\author[DoM]{Moysey Brio \corref{mycorrespondingauthor}}
\cortext[mycorrespondingauthor]{Corresponding author}
\ead{brio@math.arizona.edu}
\author[Insa]{Jean-Guy Caputo}
\author[PaM]{Kyle Gwirtz}
\author[NC]{Jinjie Liu}
\author[Mphi]{Andrei Maimistov}

\address[DoM]{Department of Mathematics, University of Arizona, Tucson, Arizona 85721 }
\address[Insa]{Laboratoire de Math\'ematiques, INSA Rouen Normandie, 76801 Saint-Etienne du Rouvray, France}
\address[PaM]{Program in Applied Mathematics, University of Arizona, Tucson, Arizona 85721 }
\address[NC]{Department of Mathematical Sciences, Delaware State University, Dover, DE 19901}
\address[Mphi]{Department of Solid State Physics and Nanostructures, \\
National Research Nuclear University, Moscow Engineering Physics Institute, \\
Kashirskoe sh. 31, Moscow, 115409 Russia}

\begin{abstract}
We combine scattering theory, Fourier, traveling wave 
and asymptotic analyses 
together with numerical simulations to present interesting and practically 
useful properties of femtosecond pulse interaction with thin films.
The dispersive material is described by a single resonance Lorentz 
model and its nonlinear extension with a cubic Duffing-type nonlinearity. 
A key feature of the Lorentz dielectric function is that its 
real part becomes negative between its zero and its pole, generating a forbidden region.
We illustrate numerically the linear interaction of the 
pulse with the film using both scattering theory and Fourier analysis. 
Outside this region we show the generation of a sequence of pulses separated 
by round trips in the Fabry-Perot cavity due to multiple reflections. 
When the pulse spectrum is inside the forbidden region, we observe total reflection. 
Near the pole of the dielectric function, we demonstrate the slowing 
down of the pulse (group velocity tending to zero) in the medium that behaves as a
high-Q cavity. We use the combination of analysis and simulations in
the linear regime to validate the delta function approximation of the thin layer; 
this collapses the forbidden region to a single resonant point of the spectrum.
We also study the single cycle pulse interaction with a thin film and show 
three distinct types of reflection: half-pulse, sinusoidal wave train and 
cosine wavelet. 
Finally we analyze the influence of a strong nonlinearity and observe that the film
switches from reflecting to trasparent.
\end{abstract}


\end{frontmatter}

\section{Introduction}
\label{sec:intro}

The interaction of femtosecond pulses with thin dispersive nonlinear films has been a subject 
of numerous recent theoretical, numerical and experimental 
studies  \cite{belous,jepsen,ckm06,mishina,Kaz:Ma:05,rosanov17}. For many transparent or weakly 
absorbent materials such as insulators, glasses, doped glasses, semiconductors and amorphous materials, 
the dielectric function can be often be described by a single resonance Lorentz model or a combination 
of such models \cite{fox, horiba,Rozanov:03,Molon:05,Trof:13}. Many
such materials have negligible or zero damping coefficient (transparent Lorentz materials) \cite{horiba}. 
One of the key features of the Lorentz dielectric function is that -between its zero and its pole- the real 
part is negative so that the wave vector is purely imaginary. This region of total reflection 
comes under different names, e.g.  forbidden zone in quantum mechanics, non-propagating region in 
electromagnetic wave theory, stop band in optical filters,  Restrahlen band in bulk solids with 
crystalline structure, and polaritonic gap in photonic crystals \cite{macleod,patterson,rung,fox}. This 
property is used in numerous applications:  thin optical film filters, 
spectroscopic ellipsometry and artificial bulk and surface meta materials 
\cite{harland, rung, ashkenazi}.

In this article, we focus on the various consequences of the presence of the forbidden region 
on the interaction of a short pulse with the thin film. For the linear Lorentz oscillator 
model and an incident pulse, an exact analytic solution is available in integral form involving 
Green's functions, inverse Fourier and Laplace transforms. However it is difficult to extract useful design 
information from these complicated formulas; this requires either asymptotic analysis or various 
simplifying assumptions \cite{gralak,bbj03}.  In our study, we  combine scattering theory, 
Fourier, traveling wave and asymptotic analyses
together with one-dimensional finite-difference time-domain (FDTD) numerical simulations 
\cite{Blokhincev,Kozlov:04,Kozlov:12} 
to provide interesting and practically useful scattering properties of thin films.  
We illustrate numerically the linear interaction of the
pulse with the film using both scattering theory and Fourier analysis.
We show in particular, the generation of a sequence of pulses separated by round trips in 
the Fabry-Perot cavity due to multiple reflections,  the total reflection and high-Q property of 
the cavity due to the slowing down (group velocity tending to zero) near the pole of the Lorentz 
dielectric function. 
The filtering property is shown for a pulse whose spectrum overlaps 
the forbidden zone. We continue this study by validating the delta function approximation of the thin layer.
We also consider the interaction of a single cycle pulse with the thin film and
show three possible types of reflection: half-pulse, sinusoidal wave train and
cosine wavelet.  Finally, a strong cubic Duffing-type polarization nonlinearity 
is studied. It demonstrates that the film
switches from being totally reflecting to being transparent. \\
The article is organized as follows. Section 2 describes the Lorentz-Duffing model. Section 3 presents the 
FDTD algorithm to solve the one-dimensional Maxwell-Lorentz-Duffing equations and the 
numerical procedure to compute the scattering coefficients. In section 4, we review the scattering
theory for the finite slab and the delta function approximation, compare the reflection coefficients
for different slab thicknesses and describe the procedure to compute the scattering coefficients
from the time series of the numerical solution. 
In section 5, we present numerical results for pulses with spectra near and within the forbidden region.
We also consider a single cycle pulse interacting with the thin film and 
the switching effect of the strong nonlinearity. 
Conclusions are presented in the final section.

\section{The model}

We consider a simplified description of the interaction of 
an electromagnetic wave with
a ferroelectric material. The Lagrangian density for the vector
potential $A$ and the polarization $P$ is 
\be \label{lag}
L = \epsilon_0 ({A_t^2 \over 2} -c^2{A_x^2 \over 2} )
+ [{\tau^2 \over \epsilon_0 }
({P_t^2 \over 2} -\alpha {P^2 \over 2}-\beta {P^4 \over 2} )
- A_t P  ] {\cal I}(x),
\ee
where we have used dimensional quantities, where the subscripts
indicate partial derivative, where $\tau$ is a characteristic time of 
the material, $\alpha$ and $\beta$ are characteristic 
parameters of the ferroelectric and where 
${\cal I}(x)$ is the indicator function
of the material; for a slab ${\cal I}(x)=1,~~{\rm if} ~~0<x<w$, else ${\cal I}(x)=0$. In \cite{ckm06}, we had written a similar density but 
used dimensionless units.

The Euler-Lagrange equations are
\begin{eqnarray} \label{AtPt}
\epsilon_0 (A_{tt} -c^2 A_{xx}) = P_t {\cal I}(x)  , \\
{\tau^2 \over \epsilon_0}( P_{tt} + \alpha P + \beta P^3) = -A_t~~,
\end{eqnarray}
where the second equation only exists in the medium. Introducing the
electric field component $E= -A_t$ results in the dimensional system
\begin{eqnarray} 
\epsilon_0 (E_{tt} -c^2 E_{xx}) = -P_{tt} {\cal I}(x), \label{Et} \\
{\tau^2 \over \epsilon_0}( P_{tt} + \alpha P + \beta P^3) = E~~,\label{Pt}
\end{eqnarray}

We normalize $E$ and $P$ as $E= E_0 e,~P=P_0 p$ and get
our final system
\begin{eqnarray} 
e_{tt} -c^2 e_{xx} = -{1 \over \xi} p_{tt} {\cal I}(x), \label{et}\\
p_{tt} + \alpha p + \beta P_0^2 p^3 = {\xi \over \tau^2} e~~,\label{pt}
\end{eqnarray}
where we have introduced the dimensionless parameter 
\be \label{xi}
\xi =  {\epsilon_0 E_0 \over P_0} .
\ee
The system of equations (\ref{et},\ref{pt}) 
describes the coupling of a wave to an oscillator. It appears in
various applications, see
Lamb's book \cite{lamb} for examples in mechanics.

\section{Finite difference time domain numerical procedure }

The system (\ref{et},\ref{pt}) is solved using a standard Yee Finite Difference 
Time Domain (FDTD) algorithm \cite{fdtd} on a staggered space-time grid for the
displacement field $D= \epsilon_0 E +P$ and the magnetic field $H$.
The one-dimensional Maxwell-Lorentz-Duffing equations are 
\begin{eqnarray} \label{maxwell}
  &&  D_t  = -H_x , \\
  &&  \mu H_t = -E_x , \\
  &&  D =  \epsilon_0 E  + P , \\
  && {\tau^2 \over \epsilon_0}( P_{tt} + \alpha P + \beta P^3) = E .
\end{eqnarray}
They are approximated using the following discretization to update in 
time $H, P, D$, and $E$ fields,  respectively, 
\begin{eqnarray} \label{fdtd}
&& \mu\frac{H^{n+\frac{1}{2}}_{j+\frac{1}{2}}-H^{n-\frac{1}{2}}_{j+\frac{1}{2}}}{\Delta t} = - \frac{E^{n}_{j+1}-E^{n}_{j}}{\Delta x} , \\
&& \frac{\tau^{2}}{\epsilon_{0}}\; \frac{P^{n+1}_{j}-2P^{n}_{j}+P^{n-1}_{j}}{\Delta t^{2}} + \tilde{\alpha} P^{n}_{j} + \tilde{\beta} (P^{n}_{j})^3 =E^{n}_{j}, \\
&& \frac{D^{n+1}_{j}-D^{n}_{j}}{\Delta t} = -\frac{H^{n+\frac{1}{2}}_{j+\frac{1}{2}} - H^{n+\frac{1}{2}}_{j-\frac{1}{2}}}{\Delta x}, \\
&& E^{n+1}_{j}=\frac{1}{\epsilon_{0}}(D^{n+1}_{j}-P^{n+1}_{j}) , \\
&& \text{where} \;\;\; \tilde{\alpha} = {\tau^2 \over \epsilon_0} \alpha, ~~~
\tilde{\beta} = {\tau^2 \over \epsilon_0} \beta.
\end{eqnarray} 
An array $ \displaystyle H^{n+\frac{1}{2}}_{j+\frac{1}{2}}$ approximates an exact magnetic field evaluated at $x=(j+\frac{1}{2}) \Delta x$ and $t=n \Delta t$. The other arrays for $P, D, $ and $E$ are interpreted similarly.

\noindent In the perfectly matched layer (PML) the $E$ and $H$ update equations are modified as follows \cite{fdtd},  
\begin{eqnarray} \label{pml}
&& \mu\frac{H^{n+\frac{1}{2}}_{j+\frac{1}{2}}-H^{n-\frac{1}{2}}_{j+\frac{1}{2}}}{\Delta t} = - \frac{E^{n}_{j+1}-E^{n}_{j}}{\Delta x} -\sigma^H_{j+\frac{1}{2}} \frac{H^{n+\frac{1}{2}}_{j+\frac{1}{2}}+H^{n-\frac{1}{2}}_{j+\frac{1}{2}}}{2} , \\
&& \epsilon \frac{E^{n+1}_{j}-E^{n}_{j}}{\Delta t} = -\frac{H^{n+\frac{1}{2}}_{j+\frac{1}{2}} - H^{n+\frac{1}{2}}_{j-\frac{1}{2}}}{\Delta x}-\sigma^E_j \frac{E^{n+1}_{j}+E^{n}_{j}}{2},
\end{eqnarray} 
with matched electric and magnetic conductivities, 
$\frac{\sigma^H(x)}{\mu}=\frac{\sigma^E(x)}{\epsilon}=\sigma(x)$, and 
cubic conductivity, $ \displaystyle \sigma (x) =\big( \frac{x}{d}\big)^3 $. 
The amplitude of the reflected wave drops gradually as the PML layer is widened.
For example, with a $10$ point wide PML layer the reflection amplitude is about $8~10^{-5}$
and drops to $5~10^{-6}$ for a PML layer with 100 points.

\subsection{Stability}

The linearized scheme above is conditionally stable for sufficiently 
small time-steps $\Delta t$.  To quantify this, we performed a Von 
Neumann stability analysis of the linearized numerical method taking
\be \label{ehpfou} \left[ \begin{array}{c}  E^{n}_{j} \\ H^{n}_{j} \\ P^{n}_{j} \\ D^{n}_{j} \end{array} \right] = \left[\begin{array}{c}
\hat{E} \\ \hat{H} \\ \hat{P} \\ \hat{D} 
\end{array} \right] e^{i\omega n\Delta t-ikj\Delta x}. \ee
The resulting dispersion relation is 
\be \label{stab_disp}
\sin^2(\frac{1}{2}\omega \Delta t) +\frac{\mu c^2 \sin^2(\frac{1}{2}\omega \Delta t)} {\frac{\tau^{2}}{\epsilon_{0}}\Big (-\frac{4 \sin^2(\frac{1}{2}\omega \Delta t)}{( \Delta t)^2)} \Big )+\tilde \alpha  } \;  = \; \Big (c \frac {\Delta t}{\Delta x} \Big )^2. 
\ee

It can be seen as a modification of the free space dispersion relation 
\be \label{stab_dispfree}
\sin^2(\frac{1}{2}\omega \Delta t) \;  = \; \Big (c \frac {\Delta t}{\Delta x} \Big )^2. 
\ee
The Courant-Friedrich-Levy (CFL) restriction in the free space, $c \frac {\Delta t}{\Delta x} \le 1, $ is replaced by the following time step restrictions under the requirement that 
$\displaystyle  \sin^2(\frac{1}{2}\omega \Delta t) \le 1, $
\be \label{stab}
 \Delta t^2 \le  \frac{\gamma - \sqrt{\gamma^2 -16 \tilde { \alpha} \mu \tau^2 c^4 (\Delta x)^2 \sin^2(\frac{1}{2}k\Delta x) }}
 {2 \tilde {\alpha} c^2 \sin^2(\frac{1}{2}k\Delta x)  },
\ee
where 
$$\displaystyle \gamma = (\tilde {\alpha}+c^2 \mu)(\Delta x)^2 
+4 c^4 \mu \tau^2 \sin^2(\frac{1}{2}k\Delta x). $$

The figure below illustrates the dependence of the time step for fixed $\Delta x$ as a 
function of the dispersion and nonlinearity parameters $\tau$ and $\alpha $ respectively. For 
sufficiently weak dispersive and nonlinear effects, the CFL restriction on the linear wave 
propagation, $\Delta t< \frac{\Delta x}{c^2}$ suffices, while for stronger dispersion and 
nonlinearities the time step has to be reduced to resolve these effects.  
\begin{figure} [H]
  \centering
  \vspace{1em}
  \resizebox{9 cm}{5 cm}{\includegraphics[angle=0]{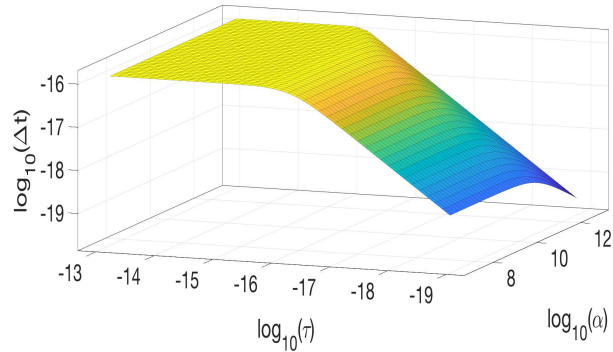}}
  \caption{\small\em Stability surface of $\Delta t$ as function of dispersion and nonlinearity parameters $\tau$ and $\alpha$.} 
  \label{Stability}
\end{figure}

\section{Scattering solution }

In the linear regime the equations (\ref{et},\ref{pt}) reduce to
\begin{eqnarray} \label{etpt_thin}
e_{tt} -c^2 e_{xx} = -{1 \over \xi} p_{tt} {\cal I}(x), \\
p_{tt} + \alpha p  = {\xi \over \tau^2} e~~,
\end{eqnarray}
where ${\cal I}(x)$ is the indicator function of the film.
Then the solution can be computed using scattering theory using a plane wave Fourier 
decomposition of the solution, an approach standard for any linear dispersive system, see for example
the scattering problem for the Schroedinger
equation described in Dodd et al \cite{dodd}. 
Writing $e,p$ in harmonic form as
$$e(x,t)= e^{i \omega t} f(x), ~ ~~  p(t)= q e^{i \omega t}$$
we get the system
\begin{eqnarray} \label{fq}
f_{xx} + k^2  f =-{k^2  \over \xi } {\cal I} q , \\
(\omega^2 -\alpha) q = -{\xi \over \tau^2} f~~,
\end{eqnarray}
which can be reduced to 
\be\label{ff}
f_{xx} + f  k^2  \left [  1 + { 1 \over \tau^2 (\alpha -c^2 k^2)}{\cal I}  
\right ] =0 .  \ee
In the slab $0<x<L$, we have
\be\label{fslab}
f_{xx} + k_0^2 f = 0,\ee
where
\be\label{k0}
k_0 = k \sqrt{ 1 + { 1 \over \tau^2 (\alpha -c^2 k^2)}} . \ee
To compute the reflection and transmission coefficients, one
writes the solutions as a left field $f^l$, middle field $f^m$ and right
field $f^r$
\begin{eqnarray}  \label{scatt}
&& f^{l} = e^{-i k x} + R e^{i k x}, \\
&& f^m = A \cos k_0 x  + B\sin k_0 x, \\
&& f^r = T e^{-i k x} .
\end{eqnarray}
At the two interfaces, $x=0,L$ the electric field and its derivative are continuous. To see this, integrate
the operator on a small interval across the interface and take the
limit of the interval going to zero. 
We then have the following interface conditions at $x=0,L$
\begin{eqnarray}
f^{l}(0)=f^m(0),~~~f^m(L)=f^r(L) ,\\
f^{l}_x (0) = f^m_x(0), ~~~f^m_x(L)=f^r_x(L).    \end{eqnarray}
This gives four linear 
equations for the four unknowns $R,T,A,B$. 
Solving for $R,T,A,B$ we get
\begin{flalign}\label{ref_trans} 
& R= {(k^2-k_0^2) \sin k_0 L \over D}, ~~~~
T= {2i k k_0 \exp i k L \over D}, && \\
& A= {2 k (-i k_0 \cos k_0 L + k \sin k_0 L) \over D},~~~~
B= {2 k ( k \cos k_0 L + i k_0 \sin k_0 L) \over D}, && \nonumber \\
& D = -2i k k_0 \cos k_0 L + (k^2+k_0^2) \sin k_0 L  . &
\end{flalign}

\subsection{Forbidden range and bound states}

\begin{figure} [H]
  \centering
  \vspace{1em}
  \resizebox{9 cm}{5 cm}{\includegraphics[angle=0]{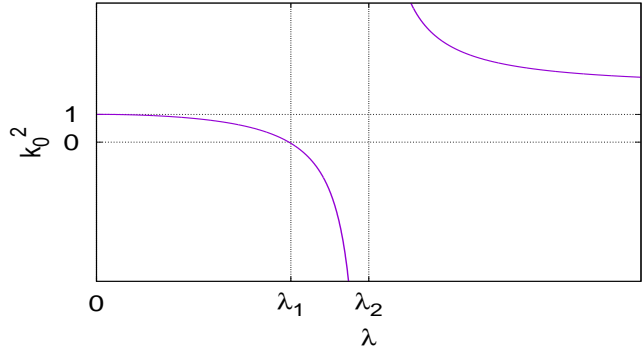}}
  \caption{\small\em Plot of $k_0^2$ as a function of the wave-length
$\lambda$. }
  \label{dis}
\end{figure}

When examining the expressions (\ref{ref_trans}) one
sees that there are special values of $k$. One of
them gives $k_0=0$ which corresponds to a pole of $R$
and $T$, the corresponding solution is called a bound state.
Another interesting $k$ is such that $k_0 \to \infty$.
These two values are shown in Fig. \ref{dis} where we
plotted $k_0$ as a function of $\lambda = 2 \pi/k$.
We indicated the value $\lambda_1 $ such that $k_0=0$ and the
value $\lambda_2$ such that $k_0 \to \infty$. These are
\begin{eqnarray}\label{l12}
&& \lambda_1 =  2 \pi {c \tau \over \sqrt{\alpha \tau^2 +1}} = 2 \pi {c \tau \over \sqrt{ {\tilde \alpha} \epsilon_0+1}} , \\
&& \lambda_2 = 2 \pi {c \over \sqrt{\alpha }} = 2 \pi {c \tau \over \sqrt{ {\tilde \alpha} \epsilon_0}} , \\
\end{eqnarray}
In the region $[\lambda_1 ;  \lambda_2]$, no propagation is possible
inside the slab as the waves are exponentially damped because $k_0$ is purely imaginary.

Bound states correspond to imaginary $k= i \kappa$; then
the field decays exponentially outside the slab. To find
them, we substitute the following ansatz into the interface boundary conditions
 \begin{eqnarray}  \label{BoundState}
 && f^{l} = e^{ \kappa x},\;\; \kappa>0 \\
 && f^m = A \cos \tilde k_0 x  + B\sin \tilde k_0 x, \\
 && f^r = T e^{- \kappa x} .
 \end{eqnarray}
The resulting solvability condition in terms of $\kappa$ is as in \cite{dodd}, 
\begin{equation}  \label{Bound_wavenumber}
\frac{-2 \kappa \tilde k_0}{\kappa^2-\tilde k_0^2}=\tan(\tilde k_0 L), 
\end{equation}
where \be\label{tilde_k0}
\tilde k_0 = \kappa \sqrt{ 1 + { 1 \over \tau^2 (\alpha +c^2 \kappa^2)}}, \ee
gives the nonlinear equation in terms of $\kappa$ for bound states allowed. Note that the  equation for $\tilde k_0$ is
exactly as in (\ref{k0}) with $k= i \kappa$.

\subsection{Thin slab : Dirac-delta function model}

In the particular case where the film thickness $L$ is small
compared to $\lambda$, we approximate 
$${\cal I} \approx {L \delta (x) } .$$
The system in harmonic component $f$ (\ref{ff}) reduces to 
\be \label{ffd}
f_{xx} + f  k^2  \left [  1 + { 1 \over \tau^2 (\alpha -c^2 k^2)}
{L \delta (x)}
\right ] =0 .
\ee
At $x=0$, we assume continuity of $f$ and have the jump condition for the first derivative of $f$ as follows,
from (\ref{etpt_thin})
\be \label{jump}
[f_x]_{0^-}^{0^+} + {k^2 f(0) L \over \tau^2 (\alpha -c^2 k^2) }= 0 .
\ee
Using these two relations together with formulas for the solution on both sides of the slab, $f^-,f^+$ from (\ref{scatt}) 
, we recover the known reflection and transmission coefficients \cite{ckm06},
\begin{eqnarray} \label{thin_ref_trans}
R= {-k L \over G}, ~~T = {2i  \tau^2 (\alpha -c^2 k^2) \over G}, \\
G = 2i  \tau^2  (\alpha -c^2 k^2) + k L
\end{eqnarray}
When $k^2 =\alpha/c^2 $, we have resonance and full reflection, $T=0$ and
$R=-1$, while for values of $k$ such that $k L$ is small the reflection is negligible and $T~1$.

\begin{figure} [H]
  \centering
  \resizebox{14 cm}{8 cm}{\includegraphics[angle=0]{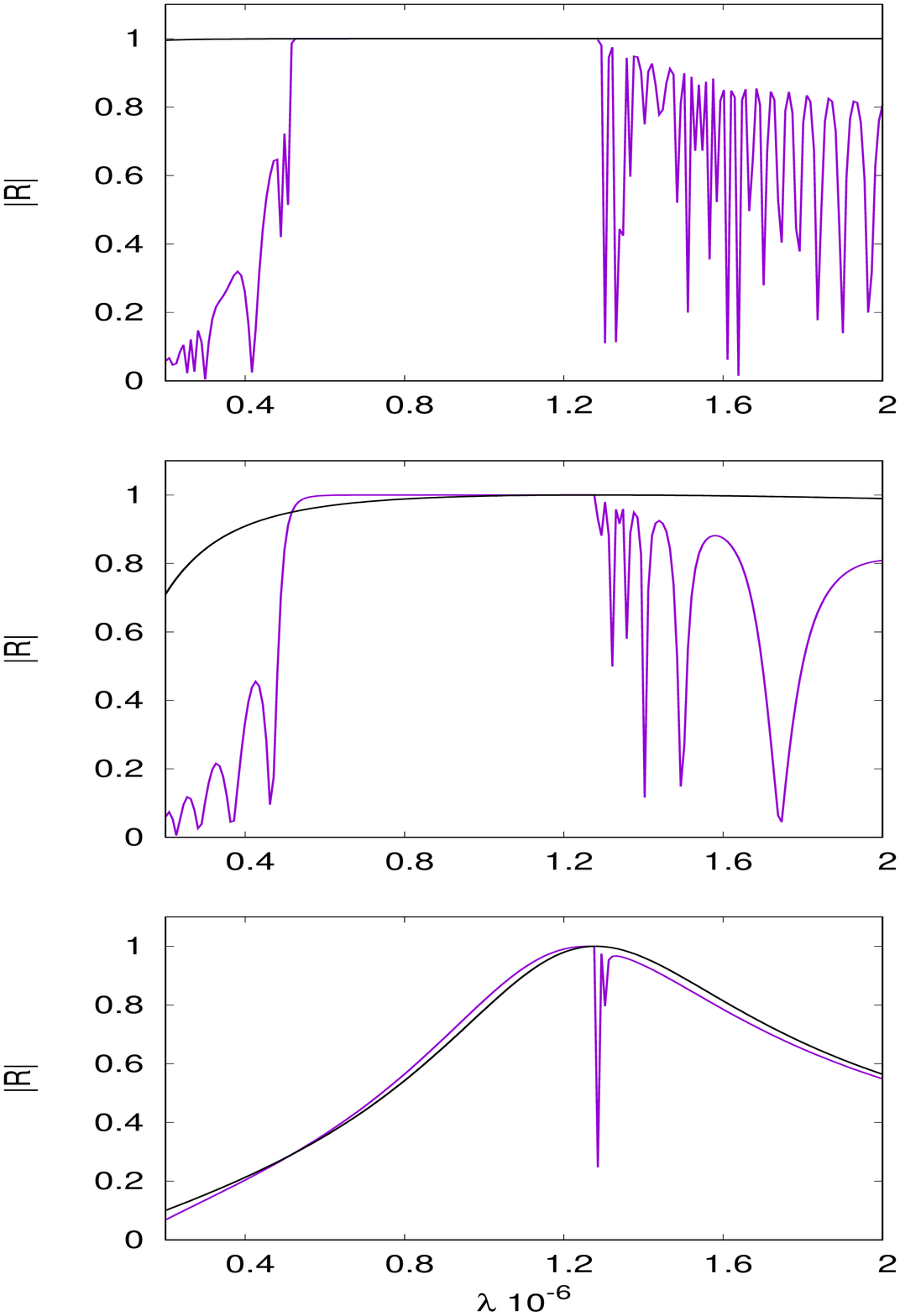}}
  \caption{\small\em Modulus of reflection coefficient $|R(\lambda)|$
for a finite slab (blue online) and the delta function approximation
(black) for widths $L = 5~10^{-6}$ (top) , $L = 5~10^{-7}$ (middle)
and $L = 5~10^{-8}$ (bottom).
}
  \label{delta_slab}
\end{figure}

To illustrate the range of validity of the delta function approximation
of the finite slab, we computed the
reflection coefficients of the delta function
(\ref{thin_ref_trans}) and of the finite slab (\ref{ref_trans})
for different thicknesses of the slab $L$. 
The results are plotted in Fig. \ref{delta_slab}.
For $L = 5~10^{-6}$, shown in the top panel of Fig. \ref{delta_slab} 
the delta function approach gives a very poor approximation of the 
reflection coefficient.
The middle panel of Fig. \ref{delta_slab} shows the case $L = 5~10^{-7}$.
Again the delta function approach fails to capture the fine features
of the reflection coefficient. Only when $L = 5~10^{-8}$, shown in the bottom
panel of Fig. \ref{delta_slab}, do the exact and delta function
approximation agree well. The delta function still fails
to predict the dip due to the forbidden range.

One can estimate the scattering coefficients directly from
the time-dependent problem (\ref{etpt_thin}).
The numerical procedure for this is described in the next section.

\subsection{Numerical computation of scattering data}

We use the following algorithm to compute reflection and transmission coefficients $R$, and $T$. 
\begin{enumerate}
\item  Fix two observation points $x=a <0$ and $x=b>L$ on each side
of the layer.

\item Run the code and record the time history of the electric field at two observation points, $E(x=a,t), ~E(x=b,t)$.

\item From the time-series $E(a,t)$ extract the incident pulse 
$E_{i}(t)$, stopping the recording before the arrival of the reflected pulse.
Then obtain the remaining record at the observation point $x=a$ extract the time-series for the reflected pulse $E_{r}(a,t)$. 

\item Take the Fourier transform (in practice the Fast Fourier
Transform (FFT)) of $E_{i}(a,t), ~E_{r}(a,t)$ and
$E_t(b,t)$. These are denoted respectively as  
${\hat E_{i}}(\omega), ~{\hat E_{r}}(\omega), ~{\hat E_{t}}(\omega)$.
\end{enumerate}
The reflection and transmission coefficients are
\be \label{RT}
R(\omega) = {{\hat E_{r}}(\omega)  \over {\hat E_{i}}(\omega) },~~~
T(\omega) e^{i\omega (b - a)}= {{\hat E_{t}} (\omega) \over {\hat E_{i}}(\omega) }, \ee
where the phase correction factor arises because the incident pulse is recorded
at location $x=a$, while the transmitted pulse is observed at $x=b$.

\section{Numerical results}

In most of our numerical experiments, the spatial domain is 
$[0;{\cal L}]$, with ${\cal L}=10^{-4}$.  The discretization was done with 
2500 uniform intervals 
except for the single cycle pulse case described later.
The time step 
$\Delta t= \Delta x / /(4 ~c)$ satisfies the stability conditions.
We chose the parameters shown in table \ref{tab1} unless stated otherwise, 
as in the case of strong nonlinearity.
\begin{table} [H]
\centering
\begin{tabular}{|c|c|c|c|c|}
   \hline
$\alpha~~(s^{-2})$ &  $L$ (m)        &  $\tau$ (s)   &  $\sigma$      & $\lambda$ (m) \\ \hline
$1.95~10^{31}$     & $5~10^{-6}$     & $3 ~10^{-16}$ &  $14~10^{-15}$ & $0.4~10^{-6}\le 1.6~10^{-6}$  \\\hline 
\end{tabular}
\vspace{20pt}
\caption{\small\em Physical parameters.}
\label{tab1}
\end{table}
For these parameters, the critical wavelengths corresponding to the
forbidden region $[\lambda_1 , \lambda_2]$ are
\be\label{l12n}
\lambda_1 = 0.517~ 10^{-6}, ~~\lambda_2=1.280~ 10^{-6}. \ee
The initial pulse propagates from the left to the right and is produced by 
the source of the following form, placed two grid points away from the PML
layer
\be\label{left_src} E(t) = e^{-(t/\sigma)^2} \sin(\omega t) . \ee

We begin our numerical examples by illustrating the pulse behavior when its 
spectrum is slightly below, within, and slightly above the forbidden 
region for three values of center wavelength $\lambda$ near the gap.

\subsection{Reflection and transmission : $\lambda=0.4 10^{-6} < \lambda_1$}

\begin{figure} [H]
  \centering
  \vspace{1em}
  \resizebox{9 cm}{4 cm}{\includegraphics[angle=0]{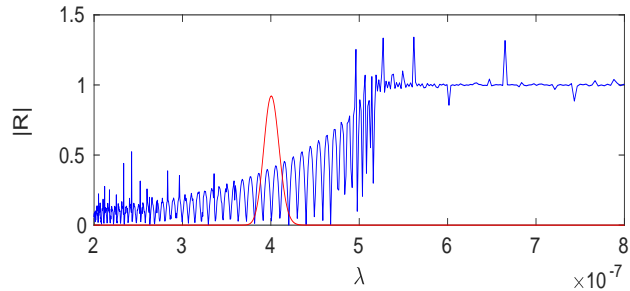}}
  \caption{\small\em Reflection coefficient 
$|R(\lambda)|$ together with $20 \times |{\hat E_i}|$. 
Same parameters as Fig. \ref{l04}. }
  \label{refl04}
\end{figure}

We first examine a pulse whose spectrum of below the forbidden region, see  Fig. \ref{refl04}.
\begin{figure} [H]
  \centering
  \resizebox{14 cm}{8 cm}{\includegraphics[angle=0]{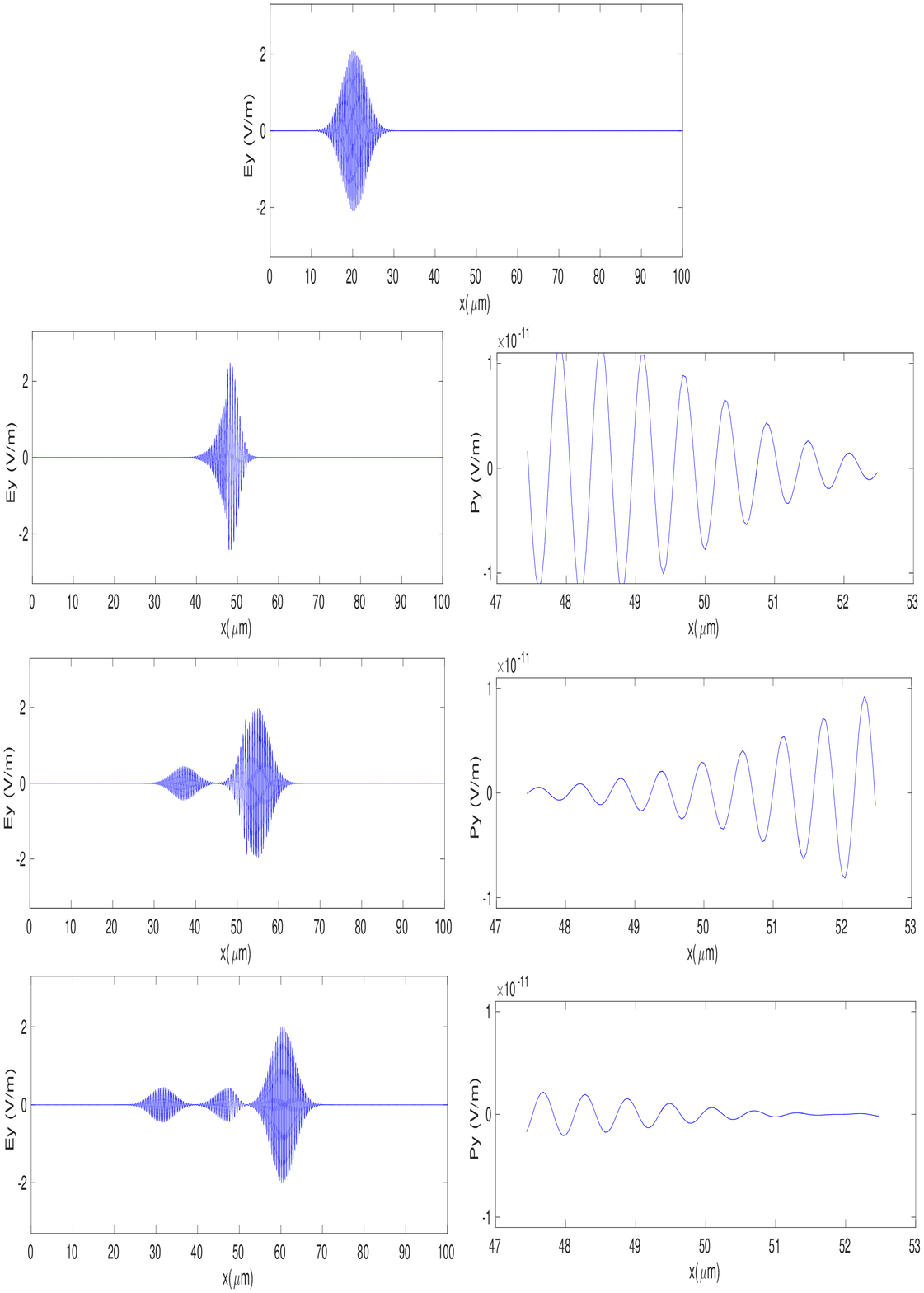}}
  \caption{\small\em Reflection / transmission of a pulse : snapshots of the 
solution $E(x,t)$ and $P(x,t)$ inside the slab for $t=10^{-13}, 2~10^{-13}, 2.3~10^{-13}$ 
and $2.5 ~10^{-13}$ (top to bottom).
}
  \label{l04}
\end{figure}
Fig. \label{l04} shows snapshots of the electric field $E(x,t)$ (left column) and polarization $P(x,t)$
(right column). The medium is located in the region $[47.5 ; 52.5]$ $\mu$m.
The first row shows the initial pulse.  
In the second row, as the pulse penetrates the slab, we see partial reflection. The polarization
is sloshing between the left and the right boundaries of the medium generating a sequence of
reflected pulses as shown in the subsequent rows. 
This dynamics is in accordance with the single frequency reflection/transmission theory 
for the Fabry-Perot cavity \cite{vaughan}.

The reflection coefficient is shown in Fig. \ref{refl04}.
Notice that even though the spectrum of the initial pulse is band
limited, we recover the full theoretical spectrum $[\pi/\Delta x , -\pi/\Delta x]$. This is because of the numerical noise induced by the round-off errors
and the discontinuity in the inital pulse on the order of the time step $\Delta t$.

\subsection{Total reflection in the forbidden range : 
$\lambda_1 < \lambda=0.6 10^{-6} < \lambda_2$}

\begin{figure} [H]
  \centering
  \vspace{1em}
  \resizebox{9 cm}{8 cm}{\includegraphics[angle=0]{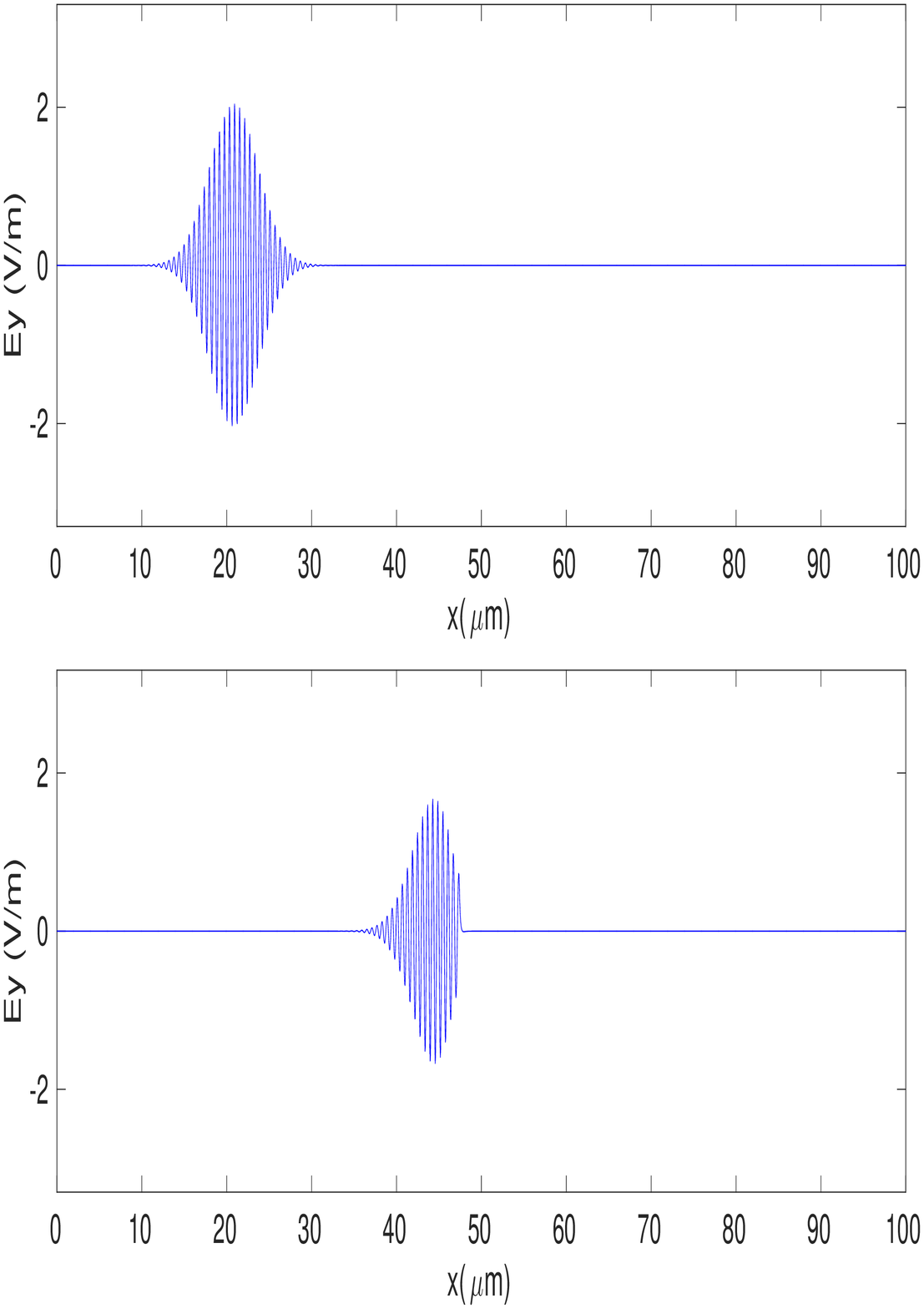}}
  \caption{\small\em Total reflection of a pulse : snapshots of the
solution $E(x,t)$ for $t=10^{-13}$ and $t=2~10^{-13}$  (top to bottom).
}
  \label{l06}
\end{figure}

\begin{figure} [H]
  \centering
  \vspace{1em}
  \resizebox{9 cm}{4 cm}{\includegraphics[angle=0]{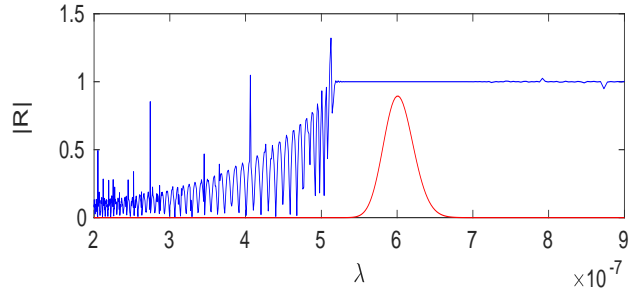}}
  \caption{\small\em Reflection coefficient
$|R(\lambda)|$ together with $20 \times |{\hat E_i}|$.
Same parameters as Fig. \ref{l06}. }
  \label{refl06}
\end{figure}

For this value of the centered wavelength $\lambda$, the pulse spectrum is in the region 
of total reflection \ref{l06} and the pulse is completely reflected
as shown in Fig. \ref{l06}. The polarization decays exponentially
inside the medium.
As expected, the reflection coefficient is equal to 1 as seen
in Fig. \ref{refl06}.
Notice again that we capture the full theoretical spectrum.

\subsection{Scattering close to bound state $\lambda_2 < \lambda=1.4 10^{-6}$ }

\begin{figure} [H]
  \centering
  \vspace{1em}
  \resizebox{16 cm}{10 cm}{\includegraphics[angle=0]{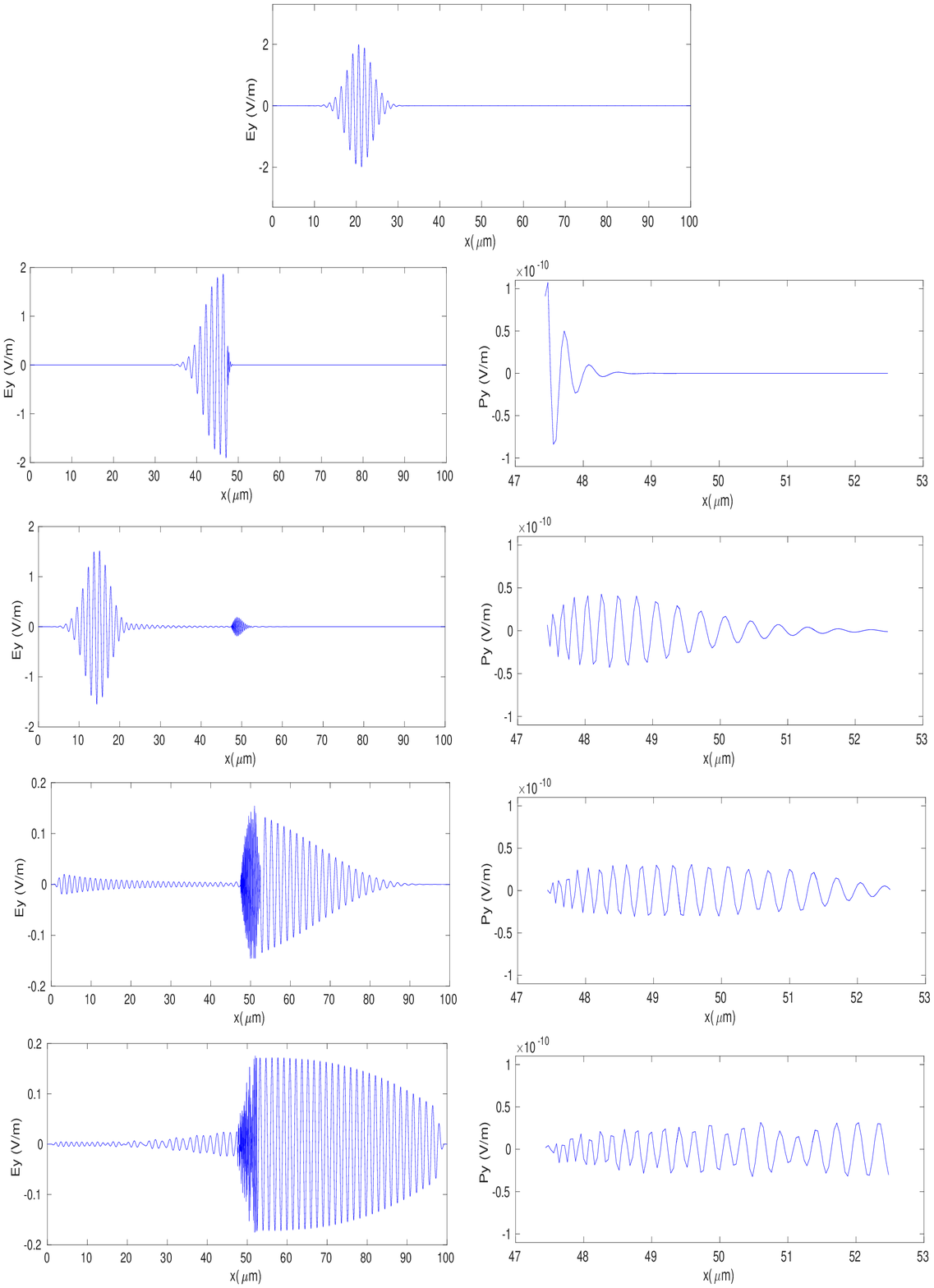}}
  \caption{\small\em Reflection / transmission of a pulse : snapshots of the
solution $E(x,t)$ and $P(x,t)$ inside the slab for $t=10^{-13}, 2~ 10^{-13}, 3~ 10^{-13}, 4~10^{-13}$
and $5 ~10^{-13}$ (top to bottom).
}
\label{l14}
\end{figure}

\begin{figure} [H]
  \centering
  \vspace{1em}
  \resizebox{12 cm}{4 cm}{\includegraphics[angle=0]{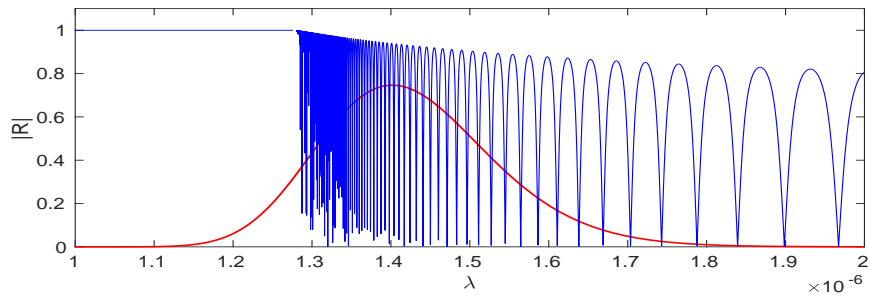}}
  \caption{\small\em Theoretical reflection coefficient $|R(\lambda)|$ 
from (\ref{ref_trans}) together with 
$20 \times |{\hat E_i}|$. The parameters are the same as in
Fig. \ref{l14}. }
  \label{refl14}
\end{figure}

For the values of $\lambda$ near the pole  $\lambda_2$, the group velocity 
$ c \sqrt(\alpha -\tau^2 c^2 k^2)/\alpha$ is near zero so that
the wave is considerably slowed down inside the slab which behaves as 
a high-Q cavity radiating long harmonic wave trains.  In Fig. \ref{l14}, the 
initial pulse is shown in the first row. The snapshots in time are 
shown in the subsequent rows for the electric field (left column) and the polarization 
inside the slab (right column). In the second row, the reflected pulse is 
leaving the computational domain and the wave inside the slab has not yet 
reached the right boundary of the slab. The film acts as a low-pass filter
due to the overlap of the pulse spectrum with the forbidden zone. 
Fig. \ref{refl14} shows the theoretical reflection coefficient, exhibiting
a singularity for $\lambda=\lambda_2$ and zeroes at $k$ such that 
$\sin k_0 L = 0$. As can be seen, the zeroes of $|R|$ accumulate near
the pole $\lambda_2$.

\subsection{Reflection of a single cycle incident pulse}

In \cite{rosanov17} it was shown that a one cycle incident pulse may 
generate a half-pulse reflected wave for a very thin film.
In this section we generalize this result by demonstrating three distinct reflection
regimes arising when a 
single cycle pulse impinges on a thin Lorentz dispersive layer.  To see this,
we consider the full Maxwell-Lorentz equations.
The exact solution of 
the one-dimensional wave equation with a source $P_{tt}$ 
is equal to the double integral of the source in space and time, \ref{Et} and assuming 
additionally a delta 
function in space leaves only time integration \cite{rosanov17}. This 
implies that the reflected wave is proportional to $P_t$ \cite{rosanov17}. Consider 
the polarization equation \ref{Pt} with a source equal to the single cycle incident 
pulse (\ref{left_src}). 
There are three distinct regimes.
\begin{enumerate}
\item  If $P_{tt}$ is the dominant term, then $P_{tt} \approx E $ so that 
$P_t$ is equal to the integral of the incident pulse,
resulting in a half-pulse reflection.

\item When the terms $P_{tt}$ and $P$ are of comparable size, the layer behaves as a 
harmonic source in time and generates a sinusoidal wave train.

\item  Finally, when $P$ is the  dominant term in the polarization equation, 
then $P \approx E $ so that $P_t$ is the time derivative of the incident 
pulse and is similar to the second derivative of a Gaussian pulse, also called a cosine wavelet.
\end{enumerate}

\begin{figure} [H]
  \centering
  \vspace{1em}
  \resizebox{12 cm}{7 cm}{\includegraphics[angle=0]{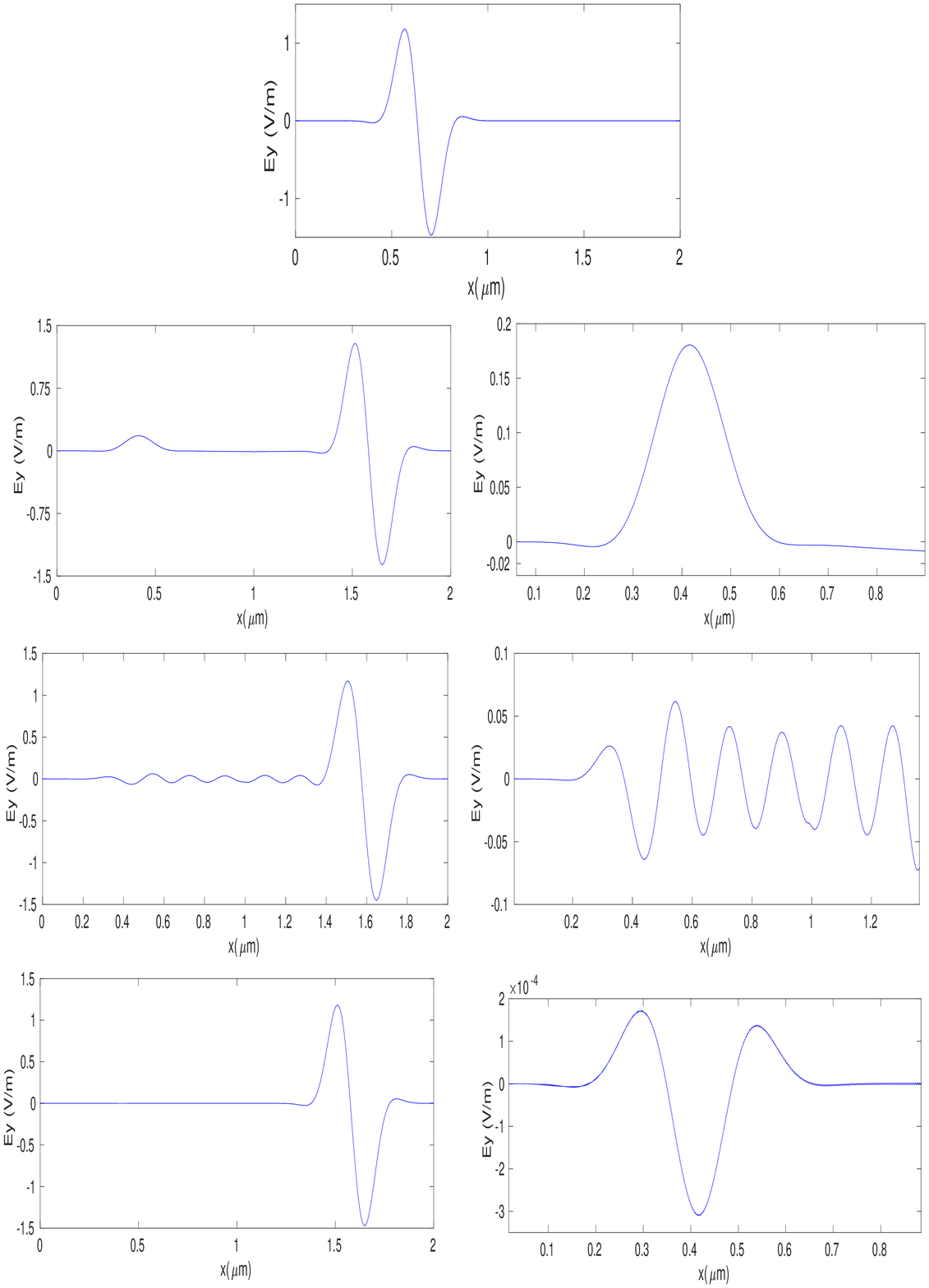}}
  \caption{\small\em Single cyle pulse interaction with a thin film.
Initial pulse (top),  transmitted and reflected pulses and blow-up
of the reflected wave for 
$\tilde \alpha, =1.156~10^9$ (2nd row),
$\tilde \alpha, =1.156~10^{12}$ (3rd row)
and $\tilde \alpha, =1.156~10^{14}$ (4th row).
The width of the initial pulse is $\sigma=0.43 ~10^{-15}$ and
the width of the layer is $L=2 ~10^{-8}$. The other parameters are the
same as in table \ref{tab1}.}
\label{eros}
\end{figure}

We solved the interaction of a single cycle incident pulse
with a thin Lorentz media layer numerically for a slab of 
thickness $L=2 ~10^{-8}$
and chose a resolution of 20 uniformly distributed points accross the slab. 
Fig. \ref{eros} illustrates the three distinct possibilities for 
the reflected wave.  The left column of Fig. \ref{eros} contains both the transmitted and 
reflected waves, while the right column shows the blow-ups of the reflected waves. Row
2 shows a half-pulse reflected wave of amplitude of about 20\% of the original pulse, this is case 1. 
Row 3 shows a sinusoidal wave train generated by 
the oscillating polarization in the film as in case 2. Finally, row 3 shows the cosine 
wavelet that has amplitude 
of about 0.01\% of the original pulse, as in case 3. It cannot be seen on the plot together with the 
transmitted wave due to the disparaty of the amplitude scales.

\subsection{Nonlinear effects}

In this subsection, we show that a strong cubic nonlinearity of the Duffing form $P^3$ added to 
the linear Lorentz model may switch the thin film from being metal-like and totally reflective to 
becoming completely transparent. We start with a numerical illustration and proceed with an 
analytic explanation of this phenomenon. Consider a strongly nonlinear medium 
described by the coefficient $\beta = 10^{40}$
so that the terms $\alpha P$ and $\beta P^3$ are of the same order. 
We choose a pulse center wavelength $\lambda = 9 ~10^{-7}$ such that the pulse spectrum
is in the forbidden band (the pulse width is as in Table \ref{tab1}). 
\begin{figure} [H]
  \centering
  \vspace{1em}
  \resizebox{9 cm}{4 cm}{\includegraphics[angle=0]{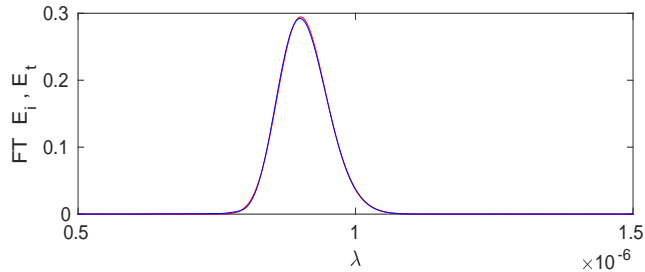}}
  \caption{\small\em Modulus of the Fourier transform of the incoming 
pulse and the transmitted pulse as a function of $\lambda$.}
  \label{nlin_fteiet}
\end{figure}
When $\beta =0$, the pulse gets reflected by the slab as described by the linear theory 
reviewed above. On the other hand, when $\beta = 10^{40}$ the pulse goes right through the 
slab with no visible reflection. Fig. \ref{nlin_fteiet} shows the Fourier transforms of both the 
incoming and transmitted fields. They are nearly indistinguishable on the plotted scale.

The heuristic explanation of this phenomena is that in the linear regime, reflection occurs due to 
destructive interference between the polarization and the field, whereas strong nonlinearity will 
change the frequency of the polarization and prevent this interference.
The medium then becomes transparent.
To justify this more rigorously, consider traveling waves for both the electric field and the 
polarization. Let $E(x-st)$ and  $P(s-st)$ and set $\epsilon =1/\beta<<1$ as
a small parameter.
Then the polarization equation  becomes 
$$(-s^2 +\alpha) \epsilon P + P^3 = \epsilon E .$$
Expanding in $\epsilon $ we obtain
$$P = (\epsilon E)^{1/3} +O(\epsilon^{2/3}). $$
For the traveling wave solutions, the wave equation reduces to
$$-s^2 E +E_{xx} = -s^2 P.$$
After substituting $P$, it becomes in the leading order $O(1)$ 
$$-s^2 E+ E_{xx}=0 $$
which is exactly the wave equation in the traveling frame outside the film.
Therefore to leading order in $\epsilon$ the incident pulse in 
not influenced by the thin film.



\section{Conclusion}

We have applied theoretical analysis and numerical simulations to present interesting 
and practically useful scattering properties of femtosecond pulses interacting with 
linear and nonlinear thin films. Combining scattering theory with numerical Fourier analysis
we obtain a consistent picture of the filtering, multiple reflection, total reflection 
and high-Q cavity regimes observed for a finite width linear film. 
We also examined the validity of the delta function approximation 
and described three possible reflection scenarios for a single cycle pulse impinging on the thin film.
Finally we presented a nonlinear switching effect.

Several observations came as a result of our study. The delta function approximation
of the medium is accurate if the central wavelength of the pulse 
is about an order of magnitude larger than the width of the layer; this
approximation shrinks the forbidden region to a single resonant wavelength. 
The generation of a half-pulse from a single cycle incident pulse is sensitive to the parameters of the 
medium. Three distinct reflections are possible, the half-pulse, the sinusoidal wave train and the 
cosine wavelet. In the absence of a dominant term in the polarization equation, a combination
of these three types of solutions will be present.
A strong nonlinearity effectively changes the refractive index of the film making
it very close to the refractive index of the outside medium. This results in a nearly perfectly
transparent nonlinear film.

\section{Acknowledgments}
M. B. thanks INSA de Rouen for an invited professorship in the spring of
2018.  The work of M.B. and J.L. were supported in part by the Air Force Office of Scientific Research under award number and FA9550-16-1-0199. J.-G. C.
was supported by the Fractal Grid project from Agence Nationale de la 
Recherche.

\end{document}